\documentclass[superscriptaddress,amsmath,amssymb,aps,pra,twocolumn,showpacs]{revtex4-1}
\usepackage{float}
\usepackage{graphicx}
\graphicspath{ {./} }
\usepackage{dcolumn}
\usepackage{tabularx}

\usepackage[dvipsnames]{xcolor}

\begin{document}

\title{Nonlinear multi-state tunneling dynamics in a spinor Bose-Einstein condensate}

\author{Z.~N. Hardesty-Shaw}
\affiliation{Department of Physics, Oklahoma State University, Stillwater, Oklahoma 74078, USA}

\author{Q. Guan}
\affiliation{Homer L. Dodge Department of Physics and Astronomy, The University of Oklahoma, Norman, Oklahoma 73019, USA}
\affiliation{Center for Quantum Research and Technology, The University of Oklahoma, Norman, Oklahoma 73019, USA}
\affiliation{Department of Physics and Astronomy, Washington State University, Pullman, WA 99164, USA}

\author{J.~O. Austin-Harris}
\affiliation{Department of Physics, Oklahoma State University, Stillwater, Oklahoma 74078, USA}

\author{D. Blume}
\affiliation{Homer L. Dodge Department of Physics and Astronomy, The University of Oklahoma, Norman, Oklahoma 73019, USA}
\affiliation{Center for Quantum Research and Technology, The University of Oklahoma, Norman, Oklahoma 73019, USA}

\author{R.~J. Lewis-Swan}
\email{lewisswan@ou.edu} \affiliation{Homer L. Dodge Department of Physics and Astronomy, The University of Oklahoma,
Norman, Oklahoma 73019, USA} \affiliation{Center for Quantum Research and Technology, The University of Oklahoma, Norman,
Oklahoma 73019, USA}

\author{Y. Liu}
\email{yingmei.liu@okstate.edu} \affiliation{Department of Physics, Oklahoma State University, Stillwater, Oklahoma 74078,
USA}

\date{\today}

\begin{abstract}
We present an experimental realization of dynamic self-trapping and non-exponential tunneling in a multi-state system
consisting of ultracold sodium spinor gases confined in moving optical lattices. Taking advantage of the fact that the
tunneling process in the sodium spinor system is resolvable over a broader dynamic energy scale than previously observed
in rubidium scalar gases, we demonstrate that the tunneling dynamics in the multi-state system strongly depends on an
interaction induced nonlinearity and is influenced by the spin degree of freedom under certain conditions. We develop a
rigorous multi-state tunneling model to describe the observed dynamics. Combined with our recent observation of
spatially-manipulated spin dynamics, these results open up prospects for alternative multi-state ramps and state transfer
protocols.
\end{abstract}

\maketitle

\section{Introduction}
\label{introduction} The phenomenon of tunneling has been widely studied in a range of physical systems including
Josephson junctions~\cite{ryu2013experimental,mullen1988combined}, superfluid annular rings~\cite{eckel2014hysteresis,
ramanathan2011superflow,fetter1967low}, waveguides~\cite{khomeriki2010nonlinear}, and Bose-Einstein condensates
(BECs)~\cite{zhang2019nonlinear,guan2020nonexponential,
Arimondo_2007,trimborn2010nonlinear,koller2016nonlinear,albiez2005direct,abbarchi2013macroscopic,chen2011many}. In each of
these examples, the tunneling description was reduced, after appropriate approximations, to a nonlinear two-state model
wherein a control parameter is ramped linearly at a rate $\alpha$ across a transition region where the two asymptotically
uncoupled states are coupled. In the absence of interactions, i.e., in a linear two-state model, the diabatic transition
between the states can be described by the linear Landau-Zener (LZ) equation, which provides the exponential probability
of tunneling between two neighboring energy levels~\cite{landau1932theorie,zener1932non}. Interactions between the
constituents of the system introduce a non-negligible nonlinearity $\gamma$ that modifies the celebrated linear LZ formula
~\cite{wu2000nonlinear,liu2002theory,zhang2019nonlinear, trimborn2010nonlinear,khomeriki2010nonlinear,guo2016nonlinear}.
Specifically, the tunneling behavior separates into three regions: (i) when $\gamma \to 0$, the dynamics are well
described by the linear LZ model; (ii)
 when $\gamma < 1$ and finite, the tunneling probability is---as for $\gamma=0$---exponential but dependent on the nonlinearity; and
(iii) when $\gamma>1$, non-exponential tunneling is observed, which is associated with dynamic self-trapping and
swallowtails~\cite{guan2020nonexponential,mumford2014impurity,albiez2005direct}.

\begin{figure}[b]
\includegraphics[width=\columnwidth,keepaspectratio]{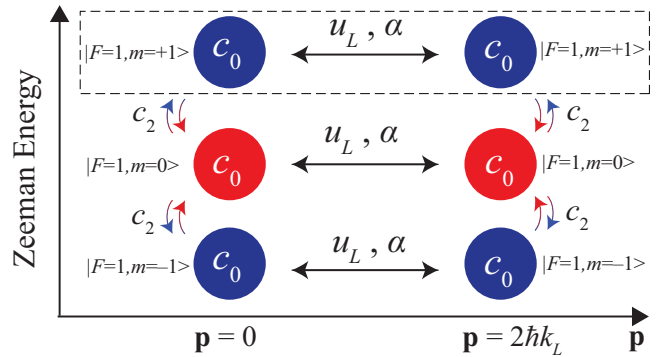}
\centering \caption{\label{fig:fig1} Schematic representation of $3 \times 2$-state tunneling model for $F$=1 spinor
gases. Circles represent atoms with momentum $\mathbf{p}$ in the $|F=1,m\rangle$ states. States with different
$\mathbf{p}$  but the same $m$ are coupled by a moving lattice, while states with different $m$ are coupled by
spin-changing interactions $c_2$. }
\end{figure}

Nonlinear multi-state tunneling has been primarily studied
theoretically~\cite{wang2006landau,fai2015multi,Sinitsyn_2017,kam2023analytical}. This paper reports an experimental
observation of non-exponential tunneling and dynamic self-trapping in a multi-state system realized by sodium spinor BECs
with multiple spin components in one-dimensional (1D) moving lattices. We demonstrate that the tunneling process in sodium
spinor BECs strongly depends on the nonlinearity induced by binary atomic interactions, and find the process is resolvable
over a broader dynamic energy scale than in prior experiments with rubidium scalar BECs~\cite{guan2020nonexponential}.
Another interesting observation is that the tunneling process in spinor gases is not always intertwined with the dynamics
of the spin degree of freedom, being spin independent for a range of conditions. This is despite the fact that appreciable
spin dynamics appear simultaneously throughout the tunneling process. These observations are well described by mean-field
Gross-Pitaevskii (GP) simulations. Our work establishes spinor BECs as a new platform for studying nonlinear multi-state
tunneling dynamics and related dynamical phase transitions in nonlinear mean-field
models~\cite{Smerzi1997selftrapping,Hazzard_2021,marino2022dynamical}.

We develop a six-state $c$-number tunneling model to provide a conceptual framework for the tunneling physics including
the spin degree of freedom. The six discrete states of our $F=1$ spinor BECs are illustrated in Fig.~\ref{fig:fig1},
wherein each spin component exhibits tunneling between $\mathbf{p}=0$ and $\mathbf{p}=2 \hbar \mathbf{k}_L$  momentum
states coupled by the moving lattice of depth $u_L$ and speed $v$ changed at a ramp rate $\alpha$ while the three spin
components are simultaneously coupled by spin-dependent interaction $c_2$. Here $\mathbf{k}_L$ is the lattice wave vector
and $\hbar$ is the reduced Planck constant. Spin-conserving interactions $c_0$ and spin-changing interactions $c_2$ both
contribute to nonlinear effects in the tunneling, with the contribution of the former being---for some atomic species such
as sodium and rubidium---notably larger than the latter. The spin degree of freedom, characterized by  $c_2$, can be
thought of as introducing a distinct second dimension that leads to an enlargement of the Hilbert space from $2$
dimensions in a spinless system (see the two states encircled by the dashed box in Fig.~\ref{fig:fig1}) to $(2F+1) \times
2$ dimensions in a spin-$F$ spinor system. Our six-state $c$-number model shows that the tunneling process in spinor BECs
is fundamentally different from the tunneling of single-component BECs even in the approximate scenario where corrections
due to spin-dependent interactions are neglected. This six-state model can be reduced in some specific circumstances to an
effective two-state $c$-number model.

The remainder of this paper is structured as follows. Sec.~\ref{experiment} introduces the experimental platform and
observations along with comparisons to standard mean-field GP simulations. Sec.~\ref{theory} introduces the theoretical
description of tunneling in a multi-state system incorporating spatial and spin degrees of freedom, which is used to
conceptually interpret our experimental observations. Finally, Sec.~\ref{conclusion} discusses applications of the
multi-state tunneling dynamics, for example on the control and coupling of spin and spatial degrees of freedom.
\vspace{-0.5pc}

\begin{figure}[t]
\includegraphics[width=\columnwidth,keepaspectratio]{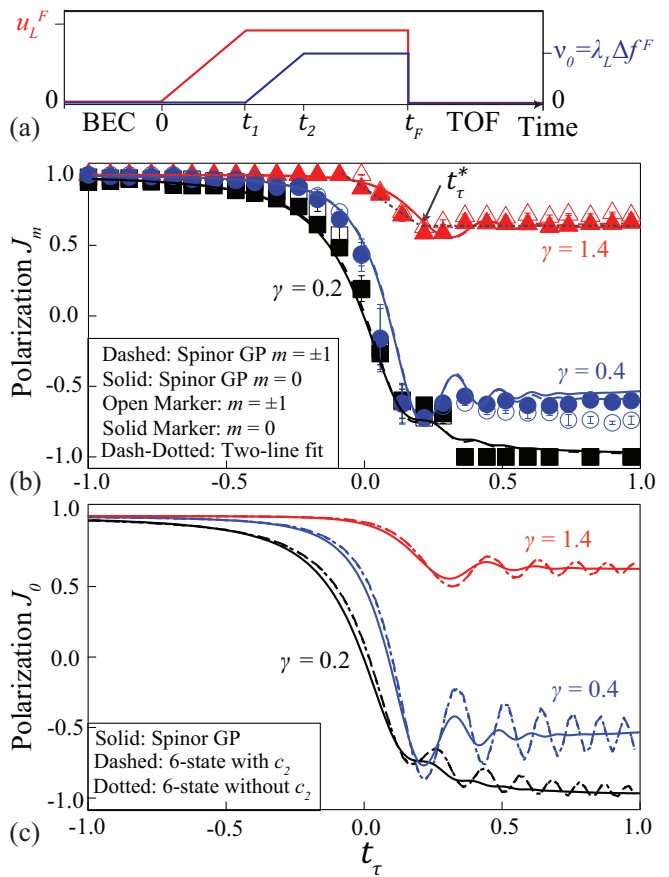}
\centering \caption{\label{fig:fig2} (a) Red (blue) lines show the experimental ramp sequence for the lattice depth
$u_L(t)$ (the lattice speed $v(t)$). The lattice reaches its maximal depth $u_L^F$ at $t_1$ while reaching its maximum
speed $v_0 = \lambda_L \Delta f^F$ at $t_2$. (b) Solid (open) markers show experimental $J_m$ versus the normalized ramp
time $t_{\tau}$ for the $|F=1,m=0\rangle$ ($|F=1,m=\pm1\rangle$) states at three different $u_L^F$ and nonlinearities
$\gamma$ with $\alpha = 4.5E_R/{\rm ms}$ and  $t_F=t_2$: $u_L^F = 0.3E_R$ corresponds to $\gamma=1.4$ with $t_\tau^* =
0.2$ (red), $u_L^F = 1.2E_R$ corresponds to $\gamma=0.4$  with $t_\tau^* = 0.1$ (blue), and $u_L^F = 2.3E_R$ corresponds
to $\gamma=0.2$ with $t_\tau^* = 0.3$ (black). Here $t_\tau^*$ is extracted via the intersection of a piece-wise linear
fit to the experimental data, as shown by the dash-dotted line for the data at $m=0$ and $\gamma=1.4$. Solid and dashed
lines show spinor GP results for the $m=0$ and $m=\pm1$ components respectively. (c) Dotted (dashed) lines show $J_0$
derived from the six-state $c$-number model with $c_2=0$ ($c_2=0.036c_0$ for our sodium system~\cite{chen2019quantum}).
These two lines are indistinguishable on the scale shown. Solid lines replot the $m = 0$ spinor GP results from panel~(b).
}
\end{figure}

\begin{figure*}[t]
\includegraphics[width=\linewidth,keepaspectratio]{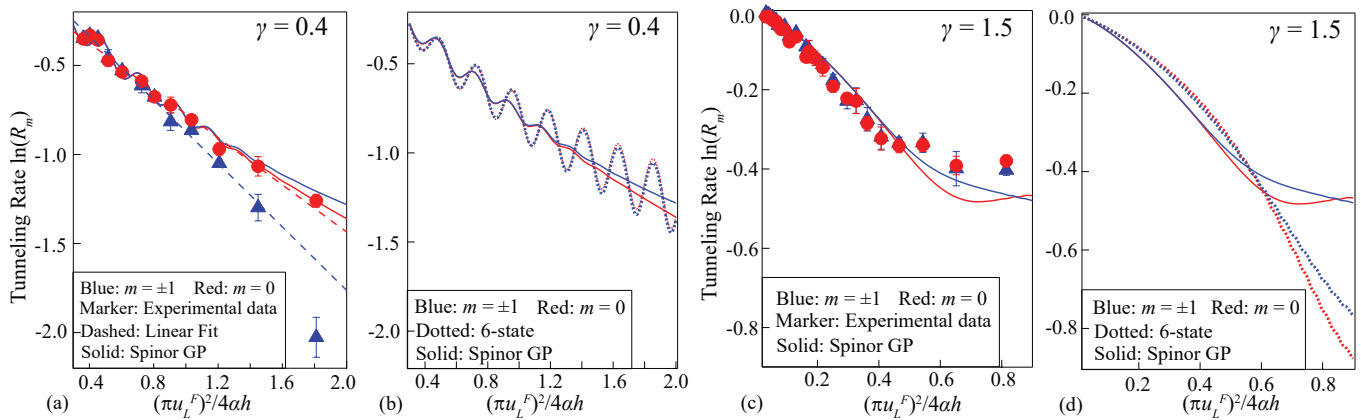}
\centering \caption{\label{fig:fig3} (a) Red (blue) color marks spin-resolved tunneling rates $\ln(R_m)$ for the
$|F=1,m=0\rangle$ ($|F=1,m=\pm1\rangle$) states versus the normalized inverse ramp rate $x=(\pi u_L^F)^2/(4 \alpha h)$ for
$t_F=t_2$ at $u_L^F=1.2E_R$ corresponding to $\gamma = 0.4$ after sequences of variable ramp rates $6.1E_R/\mathrm{ms} \le
\alpha \le 30.3E_R/\mathrm{ms}$ (scanned by setting $\Delta f^F = 9.3 E_R/h$) terminating at $t_{\tau} \gg t_{\tau}^*$.
Markers represent the experimental $\ln(R_m)$, which are fit by linear functions (see the dashed lines). Solid lines show
2D spinor GP simulation results. (b) Solid (dotted) lines represent spinor GP (six-state model) results for the
experimental conditions shown in panel (a). (c) Similar to panel (a), except with $u_L^F=0.3E_R$ corresponding to $\gamma
= 1.5$ after sequences of variable rates $1.2E_R/\mathrm{ms} \le \alpha \le 30.3E_R/\mathrm{ms}$ (scanned by setting
$\Delta f^F = 4.6 E_R/h$) terminating at $t_{\tau} = 0.2$ (which is approximately equal to $t_{\tau}^*$ for the ramp shown
in Fig.~\ref{fig:fig2}(b) for a similar value of $\gamma$, see also Appendix~\ref{app:tunnelingrates}). (d) Solid (dotted)
lines represent spinor GP (six-state model) results for the experimental conditions shown in panel (c). }
\end{figure*}

\section{Experimental results}\vspace{-0.5pc}
\label{experiment}

We construct a 1D moving optical lattice with two nearly orthogonal beams of time-dependent absolute frequency difference
$\Delta f(t)$ at wavelength $\lambda_L = 1064\,$nm. The lattice has a speed $v(t)=\lambda_L \Delta f(t)$, a depth
$u_L(t)$, and a potential $V_{\rm lat}(\mathbf{r},t) = u_L(t) \cos^2[\mathbf{k}_L \cdot \mathbf{r} - 2\pi t \Delta
f(t)/4]$. In this work, the recoil energy $E_R =h\times3.3$~kHz is much larger than the energy scales of our optical
dipole trap (ODT), the mean-field interactions, and the quadratic Zeeman shift $q=h\times 42$~Hz (see
Table~\ref{table:param} in Appendix~\ref{app:dimension}). Here $h$ is Planck's constant.

Similar to our previous
work~\cite{chen2019quantum,zhao2014dynamics,zhao2015antiferromagnetic,2023dSMA,austin2021pra,austin2021quantum}, the
experimental sequences begin with a $F=1$ spinor BEC of up to $N=1.0\times10^5$ sodium atoms in an ODT with angular
frequencies $\omega_{x,y,z}= 2 \pi \times (120,120,160)$~Hz. We apply resonant radio-frequency pulses to prepare an
initial state with fractional population $\rho_0\approx0.5$ in the $|F=1,m=0\rangle$ state and $\rho_{\pm1}\approx0.25$ in
the $|F=1,m=\pm1\rangle$ states. We then adiabatically load atoms into the lattice via a sequence shown in
Fig.~\ref{fig:fig2}(a). For $t\le t_1$, the depth $u_L$ is increased linearly from 0 to $u_L^F$ while $\Delta f$ remains
at 0 (i.e., lattices are stationary). For $t_1 \le t \le t_2$, while keeping $u_L(t)$ at $u_L^F$, we set the tunneling
control parameter $\delta(t)=-4E_R + \alpha t$ by linearly ramping $\Delta f$ at a rate $\alpha = \dfrac{h (\Delta f(t_2)
- \Delta f(t_1))}{t_2 - t_1}$ such that $\Delta f$ reaches its final value $\Delta f^F$ at $t=t_2$. Here $4E_R$ is the
kinetic energy difference of the $\mathbf{p}=0$ and $\mathbf{p}=2 \hbar \mathbf{k}_L$ momentum states.

In the absence of interactions, a fully adiabatic ramp of $\Delta f$ from $0$ to $8 E_R/h$ transfers atoms in the initial
$\mathbf{p}=0$ state to the final $\mathbf{p}=2 \hbar \mathbf{k}_L$ state. The effective coupling between these two
momentum states is maximal halfway through the ramp where $\Delta f=4E_R/h$ and $\delta = 0$. At $t=t_F$, we abruptly
switch off the lattice and ODT and let atoms ballistically expand for a certain time of flight (TOF) before monitoring
them via two-step microwave imaging~\cite{2023dSMA, chen2019quantum}.

To study tunneling dynamics in the multi-state system possessing spin and spatial degrees of freedom, we experimentally
monitor the spin-resolved polarizations
\begin{eqnarray}
\label{eq_polarization} J_m=\frac{N_{m,\mathbf{p}=0}-N_{m,\mathbf{p}\ne 0}}{N_{m,\mathbf{p}=0}+N_{m, \mathbf{p}\ne 0}}~.
\end{eqnarray}
Here $N_{m,\mathbf{p}=0}$ ($N_{m,\mathbf{p}\ne 0}$) denotes the number of atoms with zero (nonzero) momentum in the
$|F=1,m\rangle$ state. Our data indicate that the measured $N_{m, \mathbf{p}\ne 0}$ are dominated by the $\mathbf{p}=2
\hbar \mathbf{k}_L$ contribution.

Figure~\ref{fig:fig2}(b) shows measured polarizations $J_m$ of $|F=1,m\rangle$ spin components versus the normalized ramp
time $t_{\tau} = \dfrac{\alpha(t_2-t_1)}{4E_R} -1$ for a constant rate $\alpha = 4.5E_R/{\rm ms}$ at various experimental
conditions. The experimental data at each condition demonstrate that $J_m$ are very close to $1$ for negative $t_{\tau}$,
change rapidly in a narrow transition region around $t_{\tau} =0$ (where the $\mathbf{p}=0$ and $\mathbf{p}=2 \hbar
\mathbf{k}_L$ states are coupled maximally), and become approximately constant for positive $t_{\tau}$. We therefore can
extract a critical normalized ramp time $t_{\tau}^*$, via the intersection of a piece-wise linear function as illustrated
by the dash-dotted line in Fig.~\ref{fig:fig2}(b), that demarcates the end of the transition region from the final
equilibration region in which momentum state populations plateau. Interestingly, our experimental data indicate that the
observed transition regions at various experimental conditions are narrower (i.e., have smaller $t_{\tau}^* $) than those
observed previously in a rubidium system~\cite{guan2020nonexponential}. This is ascribed to the energy scales intrinsic to
the system, i.e., the larger recoil energy (see Appendix \ref{app:dimension}). Typical experimental examples are shown in
Fig.~\ref{fig:fig2}(b) for three different dimensionless spin-independent nonlinearities $\gamma =
2c_0/u_L$~\cite{wu2000nonlinear,guan2020nonexponential}. This definition of $\gamma$ ignores spin-dependent corrections as
$c_2=0.036c_0\ll c_0$ in our sodium system~\cite{chen2019quantum}, and enables direct comparisons with the nonlinear
two-state model and prior experiments on scalar BECs~\cite{guan2020nonexponential}. The dependence of the tunneling
process on $\gamma$ is rather pronounced. For example, at $\gamma=0.2$ the lattice coupling dominates and the tunneling is
spin independent resulting in the majority of the population residing in the $\mathbf{p}=0$ ($\mathbf{p}=2 \hbar
\mathbf{k}_L$) state or $J_m = 1$ ($J_m = -1$) at $t_{\tau}=-1$ ($t_{\tau}\geq t_{\tau}^*$). However, for $\gamma=0.4$ and
$1.4$ the nonlinearity has a truly non-perturbative effect with a significant fraction of atoms remaining in the
$\mathbf{p}=0$ state at the end of the ramp, i.e., the observed $J_m$ at $t_{\tau}\geq t_{\tau}^*$ plateau at a value
larger than $-1$. Such residual $\mathbf{p}=0$ populations are consistent with non-vanishing tunneling between the
asymptotically decoupled eigenstates and, for $\gamma=1.4$ self-trapping due to the presence of swallow-tails in the
adiabatic or instantaneous energy spectrum might be at play~\cite{guan2020nonexponential,wu2000nonlinear,choi_1999}.
Typically for $t_{\tau} < t_{\tau}^*$ differences between $J_0$ and $J_{\pm1}$ are small. However for $t_{\tau} >
t_{\tau}^*$, we observe statistically significant spin dependent behavior, e.g., differences between $J_0$ and $J_{\pm 1}$
curves in Fig.~\ref{fig:fig2}(b) for $\gamma=0.4$. We also conduct parameter-free numerical simulations with the
time-dependent mean-field spinor GP equation using a reduced $2$D geometry (see Appendix \ref{app:GP}) and find
theory-experiment agreement (see Fig.~\ref{fig:fig2}(b)), indicating the spinor dynamics is, as expected, in the
mean-field regime with negligible quantum and thermal effects.

We repeat the experiments for various ramp rates $\alpha$ and plot the measured spin-resolved tunneling rates,
\begin{eqnarray}
\label{eq_tunnelingRate} \ln(R_m)=\ln \left(\frac{N_{m,\mathbf{p}=0}}{N_{m,\mathbf{p}=0}+N_{m, \mathbf{p}\ne 0}} \right),
\end{eqnarray}
as a function of the dimensionless inverse ramp rate $x=(\pi u_L^F)^2/(4 \alpha h)$ for two values of $\gamma$ in
Fig.~\ref{fig:fig3}. The variable $x$ is chosen since $\ln(R_m(t))$ depends linearly on $x$ in the linear LZ two-state
model~\cite{khomeriki2010nonlinear,zhang2019nonlinear,guan2020nonexponential}. In Fig.~\ref{fig:fig3}(a), $\gamma = 0.4$
and the tunneling rate is extracted at $t_{\tau}=1.3 \gg t_{\tau}^*$. The experimental tunneling rates agree with a linear
fit (dashed lines in Fig.~\ref{fig:fig3}(a)) for $x \lesssim 1.4$ indicating exponential tunneling in this region. The
experimental results for the $|F=1,m=0\rangle$ component are also nicely reproduced by spinor GP simulations (solid lines
in Fig.~\ref{fig:fig3}), which include the density-dependent and spin-dependent interaction coefficients. Interestingly, a
splitting occurs in the tunneling rates between the $|F=1,m=0\rangle$ and $|F=1,m=\pm1\rangle$ spin components at $x > 1$.
A qualitatively similar but notably smaller splitting is also predicted by the GP simulations. Discrepancies with the
experimental observations for the $|F=1,m=\pm1\rangle$ component could be attributed to limitations of our theoretical
model, such as the reduced $2$D simulation geometry. In contrast, the data in Fig.~\ref{fig:fig3}(c), taken at a large
nonlinearity ($\gamma = 1.5$) and $t_{\tau}\approx t_{\tau}^*$, clearly shows evidence of interaction induced
non-exponential behavior in the tunneling process~\cite{guan2020nonexponential}. Another key finding from
Fig.~\ref{fig:fig3}(c) is that the observed tunneling process is spin independent. This result, when combined with the
spin-dependent behavior observed in Fig.~\ref{fig:fig3}(a), indicates a spin-dependent multi-state tunneling process that
collapses into a spin-independent process under certain conditions, such as $t_{\tau} < t_{\tau}^*$ or large $\alpha$ such
as $(\pi u_L^F)^2/(4 \alpha h) < 1$ for $\gamma = 0.4$ in Fig.~\ref{fig:fig3}(a) and $\gamma = 1.5$ in
Fig.~\ref{fig:fig3}(c). General conditions for spin-dependent tunneling related to the timescales of the system are
elucidated further in Sec.~\ref{theory}.

\begin{figure}[t]
\includegraphics[width=\columnwidth,keepaspectratio]{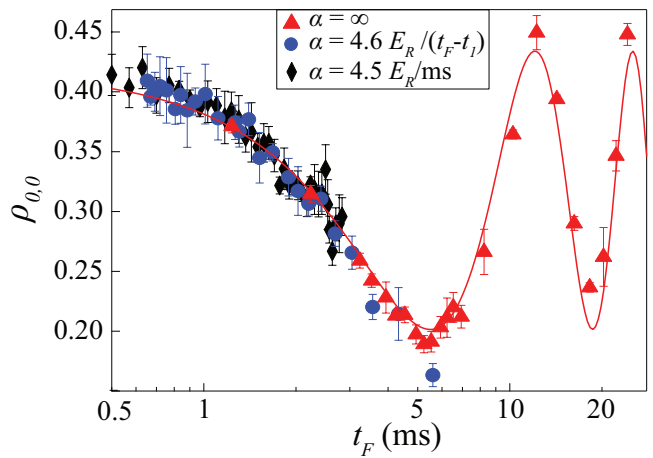}
\centering \caption{\label{fig:fig4} Experimental time evolution of the fractional spin population $\rho_{0,0}$ at $u_L^F
= 0.3E_R$ for three different ramp sequences with similar nonlinearities: a sequence with a fixed finite $\alpha$ and $t_2
= t_F$ (black; extracted from red curves in Fig.~\ref{fig:fig2}(b)), a sequence with varying $\alpha$ such that $\Delta
f^F = 4.6 E_R/h$ at $t_2 = t_F$ (blue; extracted from experimental data shown in Fig.~\ref{fig:fig3}(c)), and a quench
sequence with $\alpha = \infty$ (red; $\Delta f^F = 4.6 E_R/h$). The solid line is a sine fit to the red data for guiding
the eye.}
\end{figure}

Figure \ref{fig:fig4} displays the time evolution of $\rho_{0,0}$, the fractional population of atoms with zero momentum
in the $|F=1,m=0\rangle$ state, demonstrating that appreciable spin dynamics occur alongside the tunneling process for
three different ramp sequences with similar nonlinearities ($\gamma > 1$): black diamonds (blue circles) represent spin
oscillations extracted from experimental data shown in Fig.~\ref{fig:fig2}(b) (Fig.~\ref{fig:fig3}(c)), while red
triangles represent observations after an infinitely fast ramp, i.e., a quench with the ramp rate of $\alpha=\infty$. The
apparent agreement in spin dynamics between the three curves in Fig.~\ref{fig:fig4} is strikingly at odds with the
observation of spin-dependent tunnelling dynamics only at large moving lattice speeds and small $\alpha$ in
Figs.~\ref{fig:fig2}(b) and \ref{fig:fig3}(a). To reconcile these seemingly contradictory observations, Sec.~\ref{theory}
introduces a  six-state model that provides a rigorous framework for interpreting the experimental results.

\section{Six-state $c$-number model}
\label{theory}

As shown in Section~\ref{experiment}, the experimental results are overall well captured by the mean-field spinor GP
equation, which accounts for both density-dependent and spin-dependent interactions. Within this mean-field description
(see Appendix \ref{app:GP} for details), the dynamics are governed by $2F+1$ mean-field spinor components
$\psi_m(\mathbf{r},t)$  that depend on the spatial coordinate $\mathbf{r}$ and time $t$. These spin components are
normalized such that $\sum_{m=0,\pm1}\int|\psi_{m}(\mathbf{r},t)|^2 d \mathbf{r}=1$. This section starts with the spinor
GP equation and derives from it a six-state $c$-number model, providing a theoretical framework to interpret the observed
tunneling dynamics and to reconcile the observations of Figs.~\ref{fig:fig2}-\ref{fig:fig4}.

We start our derivation by introducing an ansatz for the spatially and time-dependent mean-field wavefunction of each
spinor component,
\begin{equation}\label{eqn:spinor_ansatz}
    \psi_m(\mathbf{r},t) = \phi_{m,0}(\mathbf{r},t) + \phi_{m,2}(\mathbf{r},t)e^{2\imath\mathbf{k}_L \cdot \mathbf{r}} .
\end{equation}
This ansatz generalizes earlier work for the single-component BECs~\cite{guan2020nonexponential,guan2020Rabi}. Since the
experiment populates predominantly two distinct momentum states $\mathbf{p}=0$ and $\mathbf{p}=2 \hbar \mathbf{k}_L$, the
ansatz accounts only for these momentum components. Consistent with the fact that the momentum width of the initial BEC is
small compared to $\hbar k_L$, we assume
\begin{equation}
    \int \left[ \phi_{m,0}(\mathbf{r},t) \right]^* \phi_{m,2}(\mathbf{r},t)e^{2\imath\mathbf{k}_L \cdot \mathbf{r}}  ~ d\mathbf{r} = 0 .
\end{equation}
Inserting Eq.~(\ref{eqn:spinor_ansatz}) into the spinor GP equation~(see Appendix \ref{app:cnumber}), with the assumption
that $\phi_{m,k}(\mathbf{r},t)$ with $k=0$ and $2$ follow a Thomas-Fermi profile and the ODT can be neglected during the
ramp protocol, the spatial dependence can be integrated out to obtain an effective time-dependent Schr\"odinger equation,
\begin{equation}\label{eqn:sixstate}
    \imath \hbar \partial_t \mathbf{d}(t) = {H}^{(6)} \mathbf{d}(t) ,
\end{equation}
where $\partial_t$ denotes the derivative with respect to time. In Eq.~(\ref{eqn:sixstate}),
$\mathbf{d}(t)=(d_{-1,0},d_{-1,2},d_{0,0},d_{0,2},d_{+1,0},d_{+1,2})^T$ is the state vector that collects the $c$-numbers
$d_{m,k}(t)$, which correspond to the $m=0,\pm1$ Zeeman and $k =0,2$ momentum components, respectively. We employ the
state normalization $\sum_{m,k} |d_{m,k}(t)|^2=1$. Since the experiment predominantly occupies the $\mathbf{p}=0$ and $2
\hbar \mathbf{k}_L$ components, the experimentally measured atom numbers $N_{m,\mathbf{p}\ne0}$ can be compared directly
with the six-state model populations $N |d_{m,2}|^2$.

The six-state $c$-number Hamiltonian ${H}^{(6)}$ can be divided into two pieces, ${H}^{(6)}= {H}^{(6,D)} + {H}^{(6,S)}$.
The Hamiltonian ${H}^{(6,D)}$ accounts for the optical lattice, the density-dependent (and spin-independent) interactions,
and the quadratic Zeeman energy $q$. The Hamiltonian ${H}^{(6,S)}$ accounts for the spin-dependent interactions. The
energy scale of $H^{(6,S)}$ is set by the spin-dependent interaction coefficient $c_2=g_2 n_{\text{mean}}$, where $g_2$
denotes the spin-dependent interaction coefficient (see Appendix \ref{app:GP}) and $n_{\text{mean}}$ the mean density.

The Hamiltonian ${H}^{(6,D)}(t)$ has the following block diagonal structure:
\begin{equation}\label{eqn:pseudo2state}
    {H}^{(6,D)}(t) =
\begin{pmatrix}
{H}^{(2)}_{-1}(t) &{0} & {0} \\
{0} & {H}^{(2)}_{0}(t) & {0} \\
{0} & {0} &  {H}^{(2)}_{+1} (t)
\end{pmatrix} ,
\end{equation}
where the $2 \times 2$ blocks ${H}^{(2)}_{m}(t)$ on the diagonal are given by
\begin{equation}\label{eqn:Hm}
   {H}^{(2)}_{m}(t)=\frac{1}{2}
    \begin{pmatrix}
     \delta(t) + 2|m|q & \frac{u_L^F}{2} + 2A(\mathbf{d},t) \\
    \frac{u_L^F}{2} + 2A^*(\mathbf{d},t) &
    - \delta(t) + 2|m|q
    \end{pmatrix} .
\end{equation}
The control parameter  $\delta(t) = -4E_R + \alpha t$ arises, as in the non-linear two-state $c$-number model for a scalar
BEC (see Ref.~\cite{guan2020nonexponential}), from the kinetic energy difference of the two coupled momentum states and
the fact that the lattice is moving. The additional $|m|q$ terms on the diagonal are due to the quadratic Zeeman shift.
The quantity
\begin{eqnarray}
A(\mathbf{d},t) = {c}_0 \sum_{m=\pm1,0}(d_{m,2}(t))^*d_{m,0}(t)
\end{eqnarray}
denotes the nonlinearity that is associated with the spin-independent interactions; here, $c_0=g_0 n_{\text{mean}}$ (see
Appendix~\ref{app:GP}). Note that $A(\mathbf{d},t)$ appears on the off-diagonals as opposed to the diagonals as in the
widely studied non-linear two-state model~\cite{wu2000nonlinear,liu2002theory,zhang2019nonlinear,
trimborn2010nonlinear,khomeriki2010nonlinear,guo2016nonlinear} (see below for further discussion). To interpret the
six-state $c$-number Hamiltonian, we first assume that $c_2$, which is about $28$ times smaller than $c_0$ for
sodium~\cite{chen2019quantum}, can be set to zero, i.e., we neglect the contribution of $H^{(6,S)}$. This assumption is,
as confirmed by  numerical simulations (see Fig.~\ref{fig:fig2}(c) and discussion below), well justified.

The Hamiltonian ${H}^{(6,D)}(t)$ has the following characteristics. (i) Even though the apparent block-diagonal structure
suggests that it decouples into three independent $2 \times 2$ blocks (i.e., a set of independent two-state models each
associated with a single Zeeman component), this is not, in general, the case since the evaluation of $A(\mathbf{d},t)$
requires---as indicated by the $\mathbf{d}$ argument---knowledge of the coefficients $d_{m,k}(t)$ of all three Zeeman
components. Even when ${H}^{(6,S)}$ is neglected, the description of tunneling in spin-1 BECs requires, in general, a
six-state model. (ii) Importantly, the nonlinearity $A(\mathbf{d},t)$ is the same in each $m$-subspace, i.e., the
tunneling dynamics in the different $m$ channels is governed by the same nonlinearity. (iii) The nonlinearity
$A(\mathbf{d},t)$ in general depends on the coherences (i.e., relative phases of the state vector elements) and not just
on population differences. Properties (i)-(iii) make the tunneling of spinor BECs in a moving optical lattice
fundamentally different from tunneling of single-component BECs in a moving optical lattice even in the approximate
scenario where corrections due to the spin-dependent interactions are neglected (i.e., in the case where $c_2$ is set to
zero). Property~(ii) also provides an explanation as to why the experimental and spinor GP tunneling data in
Figs.~\ref{fig:fig2}(b) and \ref{fig:fig3}(c) are, for a good number of parameter combinations and times, approximately
independent of $m$. Since the nonlinearity on the off-diagonals of $H^{(6,D)}$ is the same in each $m$ subspace, the spin
dependence of the tunneling dynamics should be small for certain conditions.

We now formally show that the six-state c-number model reduces for specific conditions, which are fulfilled to varying
degrees in our experiment, to an effective two-state model that neglects the spin degrees of freedom. If we define new
coefficients $b_k(t)$ through
\begin{eqnarray}
\label{eq_rescale} {b}_k(t)  = \sqrt{\frac{1}{ \vert d_{m,0}\vert^2+\vert d_{m,2}\vert^2}} d_{m,k}(t),
\end{eqnarray}
then the coefficients $b_k(t)$ are, for each $m$, except for an overall phase that does not impact the populations,
governed by the time-dependent two-state Schr\"odinger equation $\imath \hbar \partial_t \mathbf{b}(t) = {H}^{(2)}(t)
\mathbf{b}(t)$ with state vector $\mathbf{b}(t) = (b_0(t),b_2(t))^T$ and Hamiltonian $H^{(2)}(t)$,
\begin{eqnarray}
\label{eq_2state} {H}^{(2)}(t) = \frac{1}{2} \left(
\begin{array}{cc}
\delta(t) -c_0 \Delta b(t)& u_L^F/2 \\
u_L^F/2 & -\delta(t)+ c_0 \Delta b(t)
\end{array}
\right).
\end{eqnarray}
Equation~(\ref{eq_2state}) reveals that the single-particle term $\delta(t)$ is accompanied by a nonlinear detuning $c_0
\Delta b(t)$, which depends on the population difference $\Delta b(t) = \vert b_0(t) \vert^2 - \vert b_2(t) \vert^2$. This
nonlinear detuning is obtained by rewriting the original $A(\mathbf{d},t)$ term of the six-state model. Thus, provided
Eq.~(\ref{eq_rescale}) holds, the six-state model with ${H}^{(6,S)}$ set to zero formally decouples into three
independent, identical two-state models, i.e., the equations of motion and associated evolution for each Zeeman component
are identical up to a scaling factor that is associated with the (conserved) fractional population of each Zeeman
component. The Hamiltonian $H^{(2)}(t)$ is the celebrated non-linear two-state $c$-number Landau-Zener
Hamiltonian~\cite{wu2000nonlinear,liu2002theory,zhang2019nonlinear,
trimborn2010nonlinear,khomeriki2010nonlinear,guo2016nonlinear}, which was experimentally realized in
Ref.~\cite{guan2020nonexponential} using rubidium.

The inclusion of the Hamiltonian ${H}^{(6,S)}$ leads to an explicit coupling between the different $m$ channels, thereby
invalidating the above mapping to three independent $2\times 2$ Hamiltonians. However, the coupling is negligible for a
large fraction of the parameter combinations considered in Figs.~\ref{fig:fig2} and \ref{fig:fig3}. The reasons are:
First, $c_2$ is much smaller than $c_0$. Second, the length of the ramps is, in many cases, sufficiently short such that
$c_2 (t_2-t_1) / h$ is negligibly small. Third, the experimentally prepared initial state has no phase difference between
the spin components (see Appendix \ref{app:twostate}). In what follows, we use the six- and two-state c-number models to
further interpret the experimental data and spinor GP results shown in Figs.~\ref{fig:fig2}(b), \ref{fig:fig3}(a), and
\ref{fig:fig3}(c).

For the ramps shown in Fig.~\ref{fig:fig2}(c), we note the following key observations regarding the six-state model. (i)
For each of the three nonlinearities $\gamma$ considered ($\gamma = 0.2, 0.4$ and $1.4$), the polarization $J_0$ obtained
from the six-state $c$-number model with finite $c_2$ (dashed lines) and with $c_2$ artificially set to zero (dotted
lines) are essentially identical. (ii) The polarizations $J_{\pm1}$ for the $m=\pm1$ components (not plotted) are
essentially identical to the $J_0$ result shown. These two observations indicate that, within the $c$-number model, the
spin-mixing interactions of strength $c_2$ have a negligible impact on the tunneling populations for the ramps studied in
Fig.~\ref{fig:fig2}(c) (specifically, using $\alpha = 4.5E_R/$ms). Since the six-state c-number model reduces, for $c_2=0$
and the initial conditions considered in the experiment, to an effective two-state model (see above), the agreement
between the six-state models with $c_2\neq 0$ and $c_2 = 0$ shows that the dependence of $J_m$ on $\gamma$ is consistent
with what has been observed for a single-component BEC~\cite{guan2020nonexponential}.

Next, we compare the results shown in Fig.~\ref{fig:fig2}(c) for the six-state model (dashed and dotted lines) and the
spinor GP simulations (solid lines). (i) While the overall agreement between the two theories is quite good, the six-state
model predicts larger amplitude oscillations for $t_{\tau}>0$ than the spinor GP simulations. These oscillations are a
consequence of the finite ramp window, i.e., the fact that the states are not fully decoupled at
$t_{\tau}=1$~\cite{guan2020nonexponential,cao2020probing}. In-trap dynamics, which is captured by the spinor GP equation
but not by the $c$-number Hamiltonian, washes these oscillations out. (ii) For all three $\gamma$ values considered, the
polarizations obtained from the six-state model lie slightly above those obtained from the spinor GP simulations in the
transition region, where $J_0$ changes rapidly with increasing $t_{\tau}$. For example, focusing on the $\gamma = 0.2$
case (black lines), the polarization $J_0$ obtained from the GP simulations (solid black line) reaches negative values
slightly earlier (around $t_{\tau} \approx -0.05$) than those obtained from the six-state model (dashed black line). This
small discrepancy is attributable to the broadening of the resonance condition -- i.e., where the two momentum components
are maximally coupled -- in the spinor GP framework as a result of the finite momentum width or, equivalently, the
inhomogeneous spatial density profile of the condensate.

The dotted lines in Figs.~\ref{fig:fig3}(b) and \ref{fig:fig3}(d) show the six-state model tunneling rates $\ln(R_m)$ for
$\gamma = 0.4$  and $\gamma = 1.5$, respectively, as a function of the dimensionless inverse ramp rate $(\pi
u_L^F)^2/(4\alpha h)$ for the same parameters as used in  Figs.~\ref{fig:fig3}(a) and \ref{fig:fig3}(c).
Figure~\ref{fig:fig3}(b) shows reasonably good quantitative agreement between the six-state (dotted lines) and GP (solid
lines) results for ``fast'' ramps, i.e., $(\pi u_L^F)^2/(4\alpha h) \lesssim 0.8$ (here ``fast'' refers to ramp sequences
in absolute units (ms) that are comparatively short, see also Appendix~\ref{app:twostate}). Consistent with the spin
independence of the polarizations $J_m$ as discussed in the context of Fig.~\ref{fig:fig2}(c), the tunneling rates
obtained using the two theories are approximately spin-independent ($m = \pm1$ data are shown as blue lines and $m = 0$
data as red lines) for ``fast'' ramps. However, for ``slower'' ramps, $(\pi u_L^F)^2/(4\alpha h) \gtrsim 0.8$, the
$c$-number model and spinor GP results deviate for primarily two reasons. First, the $c$-number model displays large
oscillations that are centered approximately around the spinor GP predictions. These arise, as discussed previously, due
to the finite ramp window and are washed out in the GP model due to in-trap dynamics for which the timescale is set by the
spin-independent interactions of strength $c_0$ and the ODT. These are not accounted for in the $c$-number model. Second,
while the $c$-number model predicts slightly spin dependent tunneling rates for $(\pi u_L^F)^2/(4\alpha h) \gtrsim 1.5$,
the spinor GP framework yields a stronger splitting between the $m=0$ and $m=\pm1$ observables beginning from about $(\pi
u_L^F)^2/(4\alpha h) \gtrsim 1.3$. From the perspective of the six-state model, the emergence of spin dependent
observables can be understood from the fact that data in this range correspond to a $c_2 (t_2-t_1)/h$ which is no longer
completely negligible (see also Appendix \ref{app:twostate}). This emphasizes that, even though $c_2/c_0$ is small, spin
dependent processes can play a role in the observed tunneling dynamics. We interpret the fact that the deviation between
$\ln(R_0)$ and $\ln(R_{\pm1})$ is notably larger for the spinor GP than the $c$-number results as an indicator that the
precise details of the spatial dynamics of the BEC become increasingly more important as the ramp duration increases.

Comparing the $c$-number and GP predictions for $\gamma = 1.5$, Fig.~\ref{fig:fig3}(d) shows trends similar to those
observed for $\gamma=0.4$. However, as the tunneling rates are obtained much closer, at $t_{\tau}\approx t_{\tau}^*$, to
the region of maximal effective coupling, the discrepancies between the spinor GP and six-state $c$-number models are
amplified. In particular, the previously mentioned oscillations, which are a feature of the $c$-number model for finite
ramp windows, are primarily responsible for the deviations between the models, which start to emerge for $(\pi
u_L^F)^2/(4\alpha h) \approx 0.2$ and become increasingly larger as  $(\pi u_L^F)^2/(4\alpha h)$ increases.

\section{Conclusion}\vspace{-0.5pc}
\label{conclusion} We have observed non-exponential tunneling and dynamic self-trapping in $F=1$ spinor BECs that realize
a six-state $c$-number tunneling model. Our data have demonstrated that the tunneling dynamics strongly depend on the
nonlinearity induced by interactions and are resolvable over a broader dynamic energy scale than in prior experiments with
rubidium scalar BECs~\cite{guan2020nonexponential}. We have also found the tunneling process is influenced by the spin
degree of freedom under certain conditions. Our work opens up exciting prospects for alternative multi-state ramps and
state transfer protocols, including studies aimed at coupling the spatial and spin degrees of freedom. In addition, we
have introduced spinor BECs as a simulator of nonlinear multi-state quantum tunneling Hamiltonians, by utilizing the
magnetic or spin degree of freedom to enlarge the Hilbert space.

The ramps utilized in this work complement earlier sodium spinor BEC studies~\cite{2023dSMA}, in which the moving
one-dimensional optical lattice was quenched. While the quench triggered significant spatial dynamics, including
fracturing of the BEC, the spatial dynamics did not destroy the coherent spin dynamics. Together with our current work,
these studies suggest that the dynamical coupling of the spatial and spin degrees of freedom may be exploited to produce
effective two or multi-state tunneling systems dependent upon the choice of initial conditions.\vspace{-1pc}
\begin{acknowledgments}\vspace{-0.5pc}
D.~B. acknowledges support by the National Science Foundation (NSF) through grant No. PHY-2110158. R.~J. L-S. acknowledges
support by NSF through Grant No. PHY-2110052 and the Dodge Family College of Arts and Sciences at the University of
Oklahoma (OU). Z.~N.~H-S., J.~O.~A-H., and Y.~L. acknowledge support by the Noble Foundation and the NSF through Grant No.
PHY-2207777. This work used the OU Supercomputing Center for Education and Research.
\end{acknowledgments}

\appendix

\section{Gross-Pitaevskii model of spinor BEC\label{app:GP}}\vspace{-1pc}
At the mean-field level, the dynamics of a weakly-interacting $F$=1 spinor BEC subject to a time-dependent potential
$V(\mathbf{r},t)$ is modeled by the time-dependent three-component or spinor GP
equation~\cite{ueda2012review,stamper2013spinor},
\begin{widetext}
\begin{align}
\label{eqn:spinorGP} i\hbar\frac{\partial}{\partial t}\begin{pmatrix}
\psi_{-1}\\
\psi_0\\
\psi_1
\end{pmatrix}& =
\left[-\frac{\hbar^2\nabla^2}{2M_{\rm Na}} + V(\mathbf{r}, t) +
g_0(N-1)\left(|\psi_{-1}|^2+|\psi_{0}|^2+|\psi_{1}|^2\right)\right]
\begin{pmatrix}
\psi_{-1}\\
\psi_0\\
\psi_1
\end{pmatrix}
+\nonumber
\begin{pmatrix}
q & 0 & 0\\
0 & 0 & 0\\
0 & 0 & q
\end{pmatrix}
\begin{pmatrix}
\psi_{-1}\\
\psi_0\\
\psi_1
\end{pmatrix}\\\nonumber
& + g_2(N-1)\begin{pmatrix}
|\psi_{-1}|^2+|\psi_{0}|^2-|\psi_{1}|^2 & \psi_1^*\psi_0 & 0\\
\psi_1\psi_0^* & |\psi_{1}|^2+|\psi_{-1}|^2 & \psi_{-1}\psi_0^*\\
0 & \psi_{-1}^*\psi_0 & |\psi_{1}|^2+|\psi_{0}|^2-|\psi_{-1}|^2
\end{pmatrix}
\begin{pmatrix}
\psi_{-1}\\
\psi_0\\
\psi_1
\end{pmatrix} .
\\
\end{align}
\end{widetext}
Here, $\psi_m(\mathbf{r},t)$ is the mean-field GP wavefunction that is associated with the Zeeman component $m$
($m=0,\pm1$) and $M_{\text{Na}}$ denotes the mass of a $^{23}$Na atom. The two-body interactions are split into a pair of
distinct contributions:  density-density (i.e., spin-independent) collisions are characterized by the interaction
coefficient $g_0$, while spin-dependent collisions are characterized by the coefficient $g_2$. The coupling constants
$g_0$ and $g_2$ are given by $4 \pi \hbar^2/M_{\text{Na}}$, multiplied by the corresponding scattering lengths.
Specifically,
\begin{equation}
    \begin{gathered}
        g_0 = \frac{4\pi\hbar^2(a_{S=0} + 2a_{S=2})}{3 M_{\rm Na}} , \\
        g_2 = \frac{4\pi\hbar^2(a_{S=2} - a_{S=0})}{3 M_{\rm Na}} ,
    \end{gathered}
\end{equation}
where $a_{S=0}=48.9 a_0$ ($a_{S=2}=54.5 a_0$) are the $s$-wave scattering lengths for the $F=0$ ($F=2$) states ($a_0$ is
the Bohr radius)~\cite{Knoop2011PRA,chen2019quantum}. The external potential $V(\mathbf{r},t)$ contains the lattice
potential $V_{\text{lat}}(\mathbf{r},t)$ as well as the approximately harmonic ODT (see Sec.~\ref{experiment}). For the
experiments discussed in this paper, the initial $n_{\text{mean}} $ is up to $6 \times 10^{19}~\mathrm{m}^{-3}$.

The spinor GP simulation data shown in Figs.~\ref{fig:fig2}(b), \ref{fig:fig2}(c), and \ref{fig:fig3} are for $N=10^5$
atoms. We exploit the axial symmetry of the experimental system ($\omega_x=\omega_y$) and construct, following the
procedure discussed in Ref.~\cite{2023dSMA}, an effective $2$D system that accounts for the ODT, moving lattice, and
gravity. The initial state preparation mimics what is done in the experiment. We first equilibrate a single-component BEC
in the absence of the lattice, then  ``redistribute" population so that the $m=-1$, $0$, and $+1$ states have fractional
populations of $1/4$, $1/2$, and $1/4$, respectively, and finally ramp the lattice with $\Delta f=0$ to its desired value
$u_L^F$ at a rate $1.6$~$E_R/$ms. Our procedure assumes that the duration of the pulse that redistributes the population
is infinitely short; we checked that a finite pulse length does not notably change the results.

\section{$c$-number model of multi-state tunneling\label{app:cnumber}}
Section~\ref{theory} develops a six-state $c$-number model of the tunneling dynamics in a spinor BEC subject to an optical
lattice, which provides---compared to the spinor GP framework---a much simplified  description of the system. In deriving
${H}^{(6)}$, we assume that the contribution from the external ODT to the spinor GP equation can be set to zero, i.e., it
is assumed that the impact of the ODT on the spatial dynamics of the spinor BEC during the ramp can be neglected. While
this is not strictly true, the approximation is reasonable for the lattice ramps considered in Sec.~\ref{experiment},
which last up to a few ms, corresponding to at most roughly one trap oscillation period.

The frequency difference $\Delta f(t)$ of the two moving lattice beams enters via an energy splitting between the two
momentum components and takes the form,
\begin{eqnarray}
    \delta(t) =
    -4 E_R + \pi \hbar \Delta f(t) +\pi \hbar \frac{\partial \Delta f(t)}{\partial t}t.
\end{eqnarray}
For the linear ramp $\Delta f(t) = \alpha t/h$ employed in the experiment, $\delta(t)$ reduces to the equation given in
Sec.~\ref{experiment}.

The contribution $H^{(6,S)}$ to the six-state model can be written as
\begin{eqnarray}
H^{(6,S)}=
\begin{pmatrix}
S_{0} & 0 \\
0 & S_{2}
\end{pmatrix}
,
\end{eqnarray}
where the matrices $S^{(0)}$ and $S^{(2)}$ read
\begin{widetext}
\begin{eqnarray}
S_{0}=
\begin{pmatrix}{c}_2(|d_{0,0}|^2+|d_{0,2}|^2) & {c}_2(d_{0,0}d_{1,0}^*+d_{-1,2}d_{0,2}^*+2d_{0,2}d_{1,2}^*) & 0 \\
{c}_2(d_{0,0}^*d_{1,0}+d_{-1,2}^*d_{0,2}+2d_{0,2}^*d_{1,2}) & {c}_2(|d_{-1,0}|^2+|d_{1,0}|^2+|d_{-1,2}|^2+|d_{1,2}|^2) & {c}_2(d_{-1,0}d_{0,0}^*+d_{0,2}d_{1,2}^*+2d_{-1,2}d_{0,2}^*)\\
0 & {c}_2(d_{-1,0}^*d_{0,0}+d_{0,2}^*d_{1,2}+2d_{-1,2}^*d_{0,2}) & {c}_2(|d_{0,0}|^2+|d_{0,2}|^2)\nonumber
\end{pmatrix}
\\
\end{eqnarray}
\begin{eqnarray}
S_{2}=
\begin{pmatrix}{c}_2(|d_{0,0}|^2+|d_{0,2}|^2) & {c}_2(d_{0,2}d_{1,2}^*+d_{-1,0}d_{0,0}^*+2d_{0,0}d_{1,0}^*) & 0\\
{c}_2(d_{0,2}^*d_{1,2}+d_{-1,0}^*d_{0,0}+2d_{0,0}^*d_{1,0}) &  {c}_2(|d_{-1,0}|^2+|d_{1,0}|^2+|d_{-1,2}|^2+|d_{1,2}|^2) &{c}_2(d_{-1,2}d_{0,2}^*+d_{0,0}d_{1,0}^*+2d_{-1,0}d_{0,0}^*)\\
0 & {c}_2(d_{-1,2}^*d_{0,2}+d_{0,0}^*d_{1,0}+2d_{-1,0}^*d_{0,0}) & {c}_2(|d_{0,0}|^2+|d_{0,2}|^2)\nonumber
\end{pmatrix} .
\\
\end{eqnarray}
\end{widetext}
The energy scale of $H^{(6,S)}$ is set by the spin-dependent interaction coefficient $c_2$. This implies that the
contribution of $H^{(6,S)}$ is much weaker than $H^{(6,D)}$, as $c_2$ for sodium is about 28 times smaller than the
spin-independent interaction scale $c_0$~\cite{Knoop2011PRA,chen2019quantum}. However, as shown by the spin dynamics in
Fig.~\ref{fig:fig4}, we do probe time scales where the contribution from terms proportional to $c_2$ can, in principle,
give rise to non-negligible dynamics and thus it is not {\em{a priori}} clear that $H^{(6,S)}$ can be neglected.
\vspace{-1pc}
\section{Comments on units and parameters\label{app:dimension}}

The  nonlinear six-state $c$-number model is characterized by five energy scales, namely the coupling strength $u_L^F/2$,
the density-dependent interaction energy $c_0$, the recoil energy $E_R$ (which enters through the detuning), the quadratic
Zeeman shift $q$,  and the spin-dependent  interaction energy $c_2$ (see Table~\ref{table:param}). If we restrict
ourselves to situations where the six-state $c$-number model maps cleanly to the two-state $c$-number model (see
Appendices~\ref{app:cnumber} and \ref{app:twostate}), the latter two energy scales drop out of the problem: the resulting
two-state $c$-number model can be written in terms of two dimensionless energy ratios, namely $\gamma=c_0/(u_L^F/2)$ and
$(u_L^F/2)/(4E_R)$. The dimensionless non-linearity $\gamma$ is used throughout this paper to quantify the relative
strength between the density-dependent interaction $c_0$ and the coupling strength $u_L^F/2$, which is equal to the energy
gap at zero detuning for $c_0=0$. The competition between the interactions and coupling are most prominent for $\gamma$
around one~\cite{wu2000nonlinear,liu2002theory,guan2020nonexponential}; specifically, Figs.~\ref{fig:fig2}-\ref{fig:fig3}
consider $\gamma$ between $0.2$ and $1.5$.

The energy ratio $(u_L^F/2)/(4E_R)$ compares the energy gap at zero detuning with the energy gap at the beginning of the
ramp, both calculated for $c_0=0$. In the ``ideal'' nonlinear two-state Landau-Zener
model~\cite{wu2000nonlinear,liu2002theory}, where the detuning is varied from $-\infty$ to $\infty$, this energy ratio is
equal to zero, i.e., it drops out of the problem. In experimental implementations of the non-linear Landau-Zener model,
this energy ratio is finite but should be small. For our lattice based tunneling simulator results shown in
Fig.~\ref{fig:fig2}, the ratio $(u_L^F/2)/(4E_R)$ varies from about $0.3$ for $\gamma=0.2$ to $0.04$ for $\gamma=1.4$. The
smaller this energy ratio is, the more decoupled the two states are initially. The interplay of the two energy ratios
determines the width $t_{\tau}^*$ of the transition region, which we quantify empirically by performing piecewise linear
fits (see Sec.~\ref{experiment}).  It should be noted that the dimensionless transition width $t_{\tau}^*$ is expressed as
a fraction of the total ramp time $t_2-t_1$ for a given $\alpha$. As discussed in Ref.~\cite{guan2020nonexponential}, the
dimensionless non-linearity $\gamma$ and the dimensionless ramp time are normalized using ``inconsistent'' energy scales,
namely $u_L^F/2$ and $\alpha(t_2-t_1)$, respectively.

It was commented in the discussion surrounding Fig.~\ref{fig:fig2} that the width of the transition region for our sodium
tunneling simulator is narrower, if expressed in terms of the dimensionless ramp time that uses $\alpha(t_2-t_1)$ (which
depends on $E_R$) as energy unit, than that for the rubidium tunneling simulator realized by the WSU
group~\cite{guan2020nonexponential}. This can be understood by noting that their $(u_L^F/2)/(4E_R)$ values range from
$0.88$ for $\gamma=0.306$ to $0.25$ for $\gamma=1.07$. The roughly $2.5$ times larger value of $(u_L^F/2)/(4E_R)$ in the
rubidium realization compared to our sodium realization is due to the about three times larger recoil energy for the
sodium than the rubidium experiment. As a consequence, the sodium realization is closer to the ideal non-linear
Landau-Zener model that is characterized by $(u_L^F/2)/(4E_R)=0$.

\begin{table}
\begin{ruledtabular}
\begin{tabular}{ccc}
Parameter & Energy/$h$ & Role \\
\hline
$E_R$ & $3.3$~kHz & detuning in $c$-number model \\
$u^F_L$ & $1-8$~kHz& coupling in $c$-number model \\
$c_0$ & $0.7-0.9$~kHz & nonlinearity \& in-trap dynamics \\
$\hbar\omega$ & $130$~Hz & in-trap dynamics \\
$q$ & $42$~Hz & spin-mixing dynamics \\
$c_2$ & $25-32$~Hz & spin-mixing dynamics \\
\end{tabular}
\end{ruledtabular}
\caption{\label{table:param}Summary of relevant energy scales of the sodium simulator of six-state quantum tunneling. The
parameters $E_R$, $u_L^F$, $c_0$, $q$, and $c_2$ enter into the six-state $c$-number model. The trap energy scale $\hbar
\omega$ does not enter into the $c$-number six-state model but does appear in the GP formulation that incorporates in-trap
dynamics.}
\end{table}

\section{Interpretation of tunneling rates\label{app:tunnelingrates}}

Figures~\ref{fig:fig3}(c) and \ref{fig:fig3}(d) show experimental and theoretical tunneling rates for ramps that terminate
at $t_{\tau} \approx t_{\tau}^*$ and not in the middle of the second Brillouin zone, where the states are maximally
decoupled. Primarily, this is motivated by our desire to investigate a sufficiently large range of values of the
dimensionless inverse ramp rate $(\pi u_L^F)^2/(4\alpha h)$, which must be offset by technical considerations that limit
the absolute time scales over which tunneling dynamics can be well resolved in our system. For context, the dimensionless
nonlinearity $\gamma = 2c_0/u_L^F$ is tuned in our experiment by varying the lattice depth $u_L^F$. A change of the
lattice depth, in turn, changes the dimensionless inverse ramp rate $(\pi u_L^F)^2/(4\alpha h)$. Thus, to compare results
for $\gamma = 0.4$ ($u_L^F = 1.2E_R$) and $\gamma = 1.5$ ($u_L^F = 0.3E_R$) at the same value of $(\pi u_L^F)^2/(4\alpha
h)$, we must use a ramp that is by a factor of 16 longer for $\gamma=1.5$ than for $\gamma=0.4$.

\begin{figure}[t]
\includegraphics[width=\columnwidth,keepaspectratio]{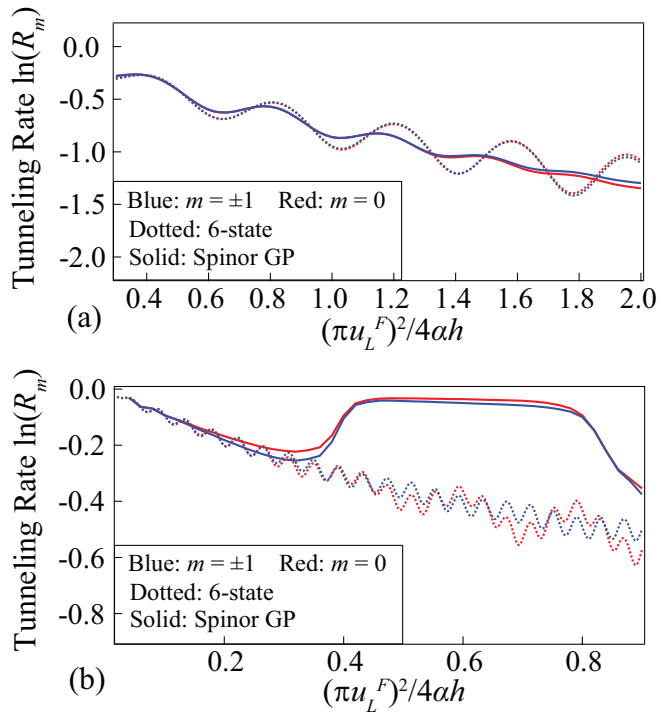}
\centering \caption{\label{fig:fig3_SM} Spin-resolved tunneling rates $\ln(R_m)$ obtained at $t_{\tau} = 1$ versus the
normalized inverse ramp rate $(\pi u_L^F)^2/(4 \alpha h)$ for (a) $\gamma = 0.4$ and (b) $\gamma = 1.5$. Solid red (blue)
lines show 2D spinor GP simulation results for the $m=0$ ($m=\pm1$) components. Dotted red (blue) lines show six-state
model results for the $m=0$ ($m=\pm1$) components. Except for $t_{\tau}$, the parameters are identical to those of
Figs.~\ref{fig:fig3}(b) and \ref{fig:fig3}(d). }
\end{figure}

To elucidate the challenges this poses (and to subsequently motivate why we use $t_{\tau} = 0.2$ for the $\gamma=1.5$
measurements shown in Fig.~\ref{fig:fig3}(c)), Figs.~\ref{fig:fig3_SM}(a) and \ref{fig:fig3_SM}(b) show the tunneling
rates $R_m$ extracted at $t_{\tau} = 1$ from spinor GP (solid lines) and six-state $c$-number (dotted lines) calculations
for $\gamma = 0.4$ and $\gamma = 1.5$, respectively. For $\gamma=0.4$, no qualitative changes are observed compared to the
results shown in Fig.~\ref{fig:fig3}(b), which uses $t_{\tau} = 1.3$. Notably, though, the spinor GP results at $t_{\tau}
= 1$ show a smaller spin dependence than those at $t_{\tau}=1.3$, suggesting that internal dynamics of the BEC (i.e.,
evolution of the spatial density profile or in-trap motion) are enhanced for larger ramp times. For $\gamma = 1.5$, in
contrast, the behavior displayed in Fig.~\ref{fig:fig3_SM}(b) (tunneling rate extracted at $t_{\tau}=1$) deviates
significantly from that in Fig.~\ref{fig:fig3}(d) (tunneling rate extracted at $t_{\tau}=0.2$, which is approximately
equal to the value of $t_{\tau}^*$ extracted in Fig.~\ref{fig:fig2}(b)). For $(\pi u_L^F)^2/(4\alpha h) \lesssim 0.3$,
which corresponds to ramps shorter than about $(t_2-t_1) \approx 2.4$~ms, the tunneling rates for the six-state model
oscillate around those for the spinor GP framework. These oscillations, and the lack thereof in the spinor GP
calculations, are understood analogously to the $\gamma = 0.4$ data presented in Figs.~\ref{fig:fig3_SM}(a) and
\ref{fig:fig3}(b). However, the spinor GP framework yields tunneling rates that abruptly upshift towards zero at $(\pi
u_L^F)^2/(4\alpha h) \approx 0.3$. This upshift, which we attribute  to non-negligible in-trap dynamics, is not captured
by the more coarse-grained six-state model.

Specifically, the finite momentum component, which gets populated by the moving lattice, undergoes significant
deceleration due to the ODT, making the distinction between the finite and the stationary $\mathbf{p} = 0$ component more
challenging for longer ramp times. When this happens, the interpretation of the dynamics within the six-state model is no
longer possible. Moreover, in this regime ($(\pi u_L^F)^2/(4\alpha h) \gtrsim 0.3$ for $\gamma = 1.5$), the experiment no
longer probes physics that can be meaningfully interpreted within the framework of tunneling physics. The data reported in
Fig.~\ref{fig:fig3}(c) are thus taken at a $t_{\tau}$ value for which the momentum components can be resolved clearly,
while maintaining $t_{\tau} \ge t_{\tau}^*$. We note that the value of $t_{\tau}^*$ is specific to our experimental
parameters; it exploits, as explained in Appendix~\ref{app:dimension}, the relative narrowness of the transition.

\section{Validity of effective two-state description\label{app:twostate}}

The discussion of Fig.~\ref{fig:fig2}(c) shows that the six-state $c$-number Hamiltonian yields essentially
spin-independent results for finite $c_2$ (using the value applicable for sodium). This behavior is accompanied, as shown
in Fig.~\ref{fig:fig3}, by tunneling rates $\ln(R_m)$ that are independent of $m$ for a good range of parameters. In both
circumstances, the spin-independent predictions of the six-state model are essentially indistinguishable from those of the
effective two-state $c$-number model (not shown).
\begin{figure}[tb]
\includegraphics[width=\columnwidth,keepaspectratio]{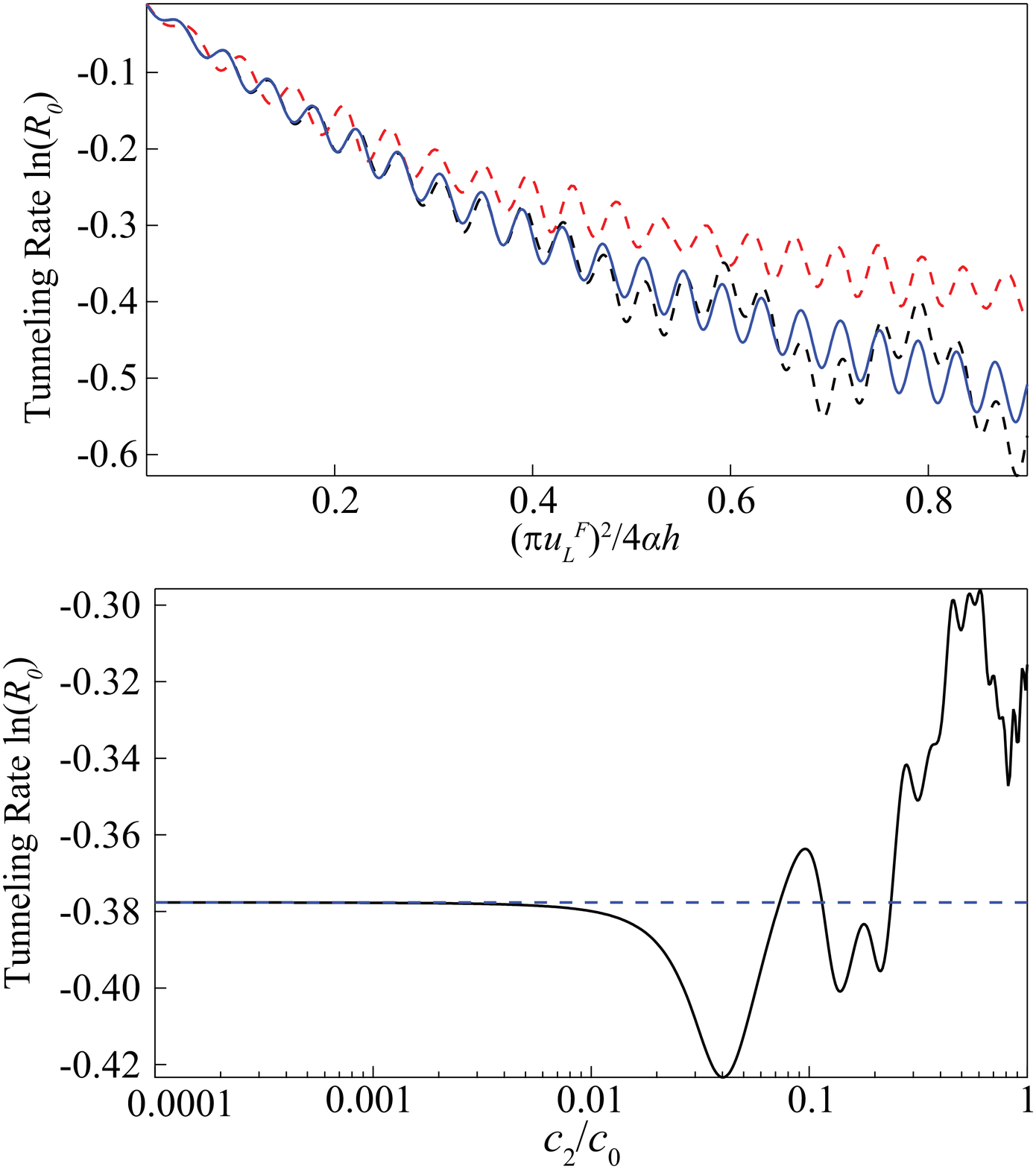}
\centering \caption{\label{fig:mdependence} (a) Spin-resolved tunneling rates $\ln(R_0)$ for the $m = 0$ component
obtained at $t_{\tau} = 1$ and $\gamma=0.4$ versus the normalized inverse ramp rate $(\pi u_L^F)^2/(4 \alpha h)$. All data
are predictions of the six-state model using $c_2/c_0 = 0$ (blue solid line), $c_2/c_0 = 0.036$ (black dashed line), and
$c_2/c_0 = 1$ (red dashed line). (b) Tunneling rate $\ln(R_0)$ obtained at fixed normalized inverse ramp rate $(\pi
u_L^F)^2/(4 \alpha h) =0.5$ as a function of the relative strength $c_2/c_0$ of the spin-dependent and spin-independent
interactions. The predictions of the six-state model are shown as the black solid line. For comparison, the horizontal
blue dashed line shows the $c_2/c_0 = 0$ result, which is equivalent to the two-state model for the chosen initial state.
Both panels use $q/h=42$~Hz. }
\end{figure}

\begin{figure}[t]
\includegraphics[width=\columnwidth,keepaspectratio]{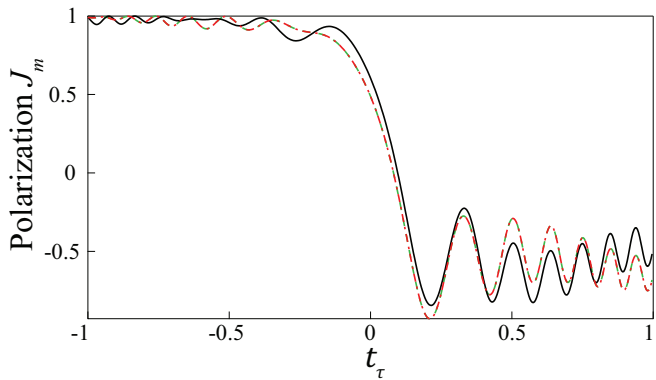}
\centering \caption{\label{fig:mdependence_part2}Polarization $J_m$, predicted by the six-state $c$-number model with
finite $c_2$, as a function of the dimensionless time $t_{\tau}$ for an initial state with spinor phases $\theta_0=0$ and
$\theta_2=-\pi$ at $t_{\tau}=-1$; this is to be contrasted with all other simulations shown in this paper, which
initialize the system using $\theta_0=\theta_2=0$ at $t_{\tau}=-1$. The black solid, green dashed, and red dash-dotted
lines show results for $m=0$, $m=-1$, and $m=+1$, respectively. The simulation parameters are $u_L^F=1.2$~$E_R$
(corresponding to $\gamma=0.4$), $q/h=42$~Hz, and $\alpha=4.5E_R$/ms. The full ramp from $t_{\tau}=-1$ to $t_{\tau}=1$
takes $1.8$~ms. The dimensionless inverse ramp rate $(\pi u_L^F)^2/(4 \alpha h)$ is equal to $2.6$. }
\end{figure}

Figure~\ref{fig:mdependence} investigates how the breakdown of the mapping between the six- and two-state models emerges
as the relative contribution of the spin-dependent interactions, characterized by $c_2/c_0$, is varied.
Figure~\ref{fig:mdependence}(a) shows the six-state model tunneling rate $\ln(R_0)$  with: (i) spin-dependent interactions
turned off, i.e.,  $c_2/c_0 = 0$ (blue solid line); (ii) $c_2/c_0 = 0.036$ (black dashed line); and (iii) $c_2/c_0 = 1$
(red dashed line). In all three cases, the spin-independent nonlinearity $\gamma$ is set to $0.4$. Since the initial state
does not contain a phase factor, the $c_2/c_0=0$ curve coincides with that for the two-state model (not shown). The value
$c_2/c_0=0.036$ describes the sodium system. In the $(\pi u_L^F)^2/(4\alpha h) \to 0$ limit, all three tunneling rate
curves approach---as they should---zero. As the ramp rate is reduced (i.e., $(\pi u_L^F)^2/(4\alpha h)$ is increased), the
$c_2/c_0 = 0$ and $0.036$ curves are nearly indistinguishable for $(\pi u_L^F)^2/(4\alpha h) \lesssim 0.3$. In this
regime, the ramps are shorter than about $(t_2-t_1) \approx 0.2$~ms on an absolute scale, so that $c_2 (t_2-t_1)/ h$ is
very small; correspondingly, the mixing of the different Zeeman states by spin-dependent interactions is expected to be
very weak and insufficient to introduce an appreciable spin-dependence into the tunneling dynamics. As $(\pi
u_L^F)^2/(4\alpha h)$  increases further, the tunneling rates for $c_2/c_0 = 0$ and $0.036$ start to deviate. The
tunneling rate for $c_2/c_0 = 1$ agrees quite well, except for a phase shift, with those for the smaller $c_2/c_0$ for
$(\pi u_L^F)^2/(4\alpha h) \lesssim 0.2$ but breaks away from those for the smaller $c_2/c_0$ for $(\pi u_L^F)^2/(4\alpha
h) \gtrsim 0.2$. Averaging over the oscillations can be seen to yield non-exponential behavior for $c_2/c_0 \approx 1$ at
$\gamma=0.4$.

Figure \ref{fig:mdependence}(b) shows the $m=0$ tunneling rate $\mathrm{ln}(R_0)$, calculated at $t_{\tau}=1$, as a
function of $c_2/c_0$ for $(\pi u_L^F)^2/(4\alpha h) = 0.5$. For $c_2/c_0 \lesssim 0.02$, the six-state model tunneling
rates (black solid line) remain very close to the $c_2/c_0 = 0$ result (blue dashed horizontal line), which coincides with
the two-state model tunneling rates. Increasing the spin-dependent interactions beyond  $c_2/c_0 \approx 0.02$ leads to
deviations between the two- and six-state model tunneling rates, consistent with the fact that the mapping from the six-
to the two-state model is expected to lose its meaning as $c_2 (t_2-t_1)/h$ or $c_2/c_0$ become non-negligible.

Section~\ref{theory} points out that the two-state description assumes a specific initial state. To demonstrate this
explicitly, Fig.~\ref{fig:mdependence_part2} shows six-state model results for  $J_m$ as a function of the normalized ramp
time $t_{\tau}$ for an  initial state that differs from what we have been using up to now, namely for a state that is
characterized by a non-vanishing spinor phase $\theta_2=-\pi$ and vanishing spinor phase $\theta_0=0$, where  $\theta_k =
\alpha_{-1,k}+\alpha_{1,k}-2\alpha_{0,k}$ and $d_{m,k}(t)$ is equal to $|d_{m,k}(t)|\exp(\imath \alpha_{m,k}(t))$. To
allow for the phase imprinting, the initial state preparation follows a modified protocol: Starting with a state
characterized by $d_{0,0}=1$ and all other $d_{m,k}=0$, the increase of the lattice depth is simulated. Population is
redistributed and the finite phase $\theta_2$ is imprinted after the lattice depth has reached its final value such that
$|d_{0,0}|^2+|d_{0,2}|^2=1/2$, $|d_{\pm1,0}|^2+|d_{\pm1,2}|^2=1/4$, $\theta_0=0$, and $\theta_2=-\pi$.
Figure~\ref{fig:mdependence_part2} shows that the non-zero value of $c_2$ leads to a spin dependence of the polarizations
$J_m$; specifically, the solid black line for $m=0$ differs from the green dashed and red dash-dotted lines for $m=\pm1$.
This behavior cannot be captured by the effective two-state $c$-number model. We emphasize that the non-vanishing spinor
phase $\theta_2$ is critical for observing the spin dependence.

The breakdown of the mapping between the six- and two-state models can also be driven by factors that are beyond the scope
of a $c$-number model. For example, evolution of the (inhomogeneous) density profile of the condensate, in-trap dynamics,
and the occupation of higher momentum states (i.e., outside the $\mathbf{p} = 0$ and $2\hbar\mathbf{k}_L$ components we
consider). This is consistent with observations throughout the main text and appendices wherein spin dependent tunneling
dynamics is more readily observed in spinor GP predictions as opposed to those obtained from the $c$-number model.

\end{document}